\documentclass[aps,prl, twocolumn]{revtex4-1}

\usepackage{geometry}                % See geometry.pdf to learn the layout options. There are lots.
\usepackage{graphicx}
\usepackage{amssymb}
\usepackage{epstopdf}
\usepackage{mathrsfs}
\usepackage{amsmath}

\begin{document}
\title{Space group symmetries and low lying excitations of many-body systems at integer fillings}
\author{Rahul Roy}
\affiliation{Department of Physics and Astronomy, University of California, Los Angeles}

\begin{abstract}
   We show that many-body systems with conserved particle number which have the symmetries corresponding to a nonsymmorphic space group have  low lying excitations for certain integer values of the particle number per unit cell. These results may be interpreted as a generalization of some band touching theorems to interacting systems. 
 \end{abstract}

\maketitle

%\date{}                                           % Activate to display a given date or no date

%\maketitle
%\section{}
%\subsection{}
\def \barph{{\overline\varphi}}
\def \ket #1{{\vert #1\rangle}}
\def \bra #1{{\langle #1\vert}}
\def \brak #1#2{{\langle#1\vert#2\rangle}}
\def\eval #1#2#3{{\langle#1\vert#2\vert#3\rangle}} 
\def\vev #1{{\langle #1\rangle}}
\def \k {\bm{k}}
\def \g {\tilde{g}}
\def \qq {\bm{q}}
\def \rr {\bm{r}}
\def \ad {a^{\dagger}} 
\def \xxp {\hat{x}_P}   
\def \yyp {\hat{y}_P}
\def \zz {\hat{z}}
\def \xx {\hat{x}}
\def \yy {\hat{y}}

\def \k {\mathbf{k}}
\def \r {\mathbf{r}}
\def \R {\bm{R}}
\newcommand {\brho}{\bar{\rho}^{g}}
\newcommand {\q} {\mathbf{q}}
\newcommand {\Q} {\mathbf{Q}}

\def\br{\mathbf{r}}
\def\bk{\mathbf{k}}
\def\bq{\mathbf{q}}
\def\bp{\mathbf{p}}
\def\bR{\mathbf{R}}
\def\bQ{\mathbf{Q}}

\def\Winf{W_\infty}
\def\vm{{\vec m}}
\def\vh{{\vec h}}
\def\ve{{\vec e}}
\def\vM{{\vec M}}
\def\nhat{{\hat n}}
\def\ehat{{\hat en}}
\def\ve{{\vec e}}
\def\ehat{\hat{\mathbf{e}}}
\def\xhat{\hat{\mathbf{x}}}
\def\yhat{\hat{\mathbf{y}}}
\def\zhat{\hat{\mathbf{z}}}

\def\grad{ \mbox{grad}}
\def\curl{ \mbox{curl}}
\def\div{ \mbox{div}}
\def\U{\ensuremath {\cal U}}
\def\S{\ensuremath {\cal S}}
\def\V{\ensuremath {\cal V}}
\def\R{\ensuremath {\cal R}}
\def\tr{\ensuremath {\mbox{tr}}}

\def \b {\mathbf{b}}
  \def \a {\mathbf{a}}
  \def \q  {\mathbf{q}}
  \def \k  {\mathbf{k}}
   \def \x  {\mathbf{x}}
   \def \K  {\mathbf{K}}
\def \btau {\mathbf{\tau}}

\def \N {\mathbf{N}}
\def \mV {\mathcal{V}}

There are relatively few results for non-interacting systems that also
hold in the presence of interactions. Kohn's theorem~\cite{kohn1961cyclotron} and Luttinger's
theorem~\cite{Luttinger:PhysRev:1960} are two celebrated examples. More recent additions to this
list include various extensions of the Lieb-Schultz-Mattis theorem as
applied to number conserving systems of fermions or bosons~\cite{oshikawa2000commensurability,hastings2004lieb}. In the
current work, we establish a technique to extend band theoretical
results to interacting systems and use it to derive interacting
counterparts to two well known band touching results for nonsymmorphic
crystals. The first band theory result asserts
that in systems with certain nonsymmorphic space groups, there are
specific wavevectors at which the electronic eigenstates must
necessarily be degenerate~\cite{konig1997electronic}; the second that for all but two of the
nonsymmorphic space groups, there are certain directions in reciprocal
space along which any energy band of the crystal must necessarily
touch another band~\cite{herring1942character,PhysRevB.59.5998}.  Together, these results apply to all the 157
nonsymmorphic space groups, which account for nearly two thirds of the
total number of space groups.

 One way to generalize the notion of a Brillouin zone to many-particle systems with interactions is to consider the phase space associated with inserting solenoidal fluxes through the holes of a periodic system.  Two Hamiltonians  which differ only by a flux insertion of multiples of $2 \pi$ through the holes of the torus are equivalent up to gauge transformations. From this perspective, the phase space of flux insertions has properties similar to the space of wavevectors for the single electron problem. 
Since  flux insertions do not destroy lattice translational invariance, at any point in this phase space, the energy eigenstates of a periodic Hamiltonian may be labeled by their crystal momenta.  The energies and the  eigenstates associated with a particular crystal momentum must form  continuous functions of the phase space variables, which may therefore be regarded as bands. We show that for a system with a  nonsymmorphic space group symmetry and  a conserved particle number and filling fractions (the number of particles per unit cell)  equal to certain integers, these bands necessarily touch~\footnote{We will also need to impose certain restrictions on the size, a feature that is shared by proofs of the higher dimensional extensions of the Lieb Schultz Mattis theorem~\cite{oshikawa2000commensurability,hastings2004lieb}.}.  

The key implication of these many-body band touching results is the existence of low lying excitations in the limit of large system sizes for integer fillings and in the presence of nonsymmorphic space group symmetries. We can establish a lower bound on the number of these excitations, which depends on the filling fraction and the nonsymmorphic space group.

The arguments that we are about to present are quite general and can be extended to a variety of systems, some of which will be discussed at a later stage.  It is simplest, however, to first consider a system of spinless fermions or bosons  in 3d with a conserved particle number which is invariant under translations by the primitive lattice vectors, $\a_{1}, \a_{2}$ and $\a_{3}$. We will also assume that the filling fraction is some integer $\nu$.  We impose  periodic boundary conditions $ (T_{\a_1})^{N}= (T_{\a_2})^N= (T_{\a_3})^{N} =1 $ for some integer, $N$, where $T_{\r}$ is the operator that corresponds to a lattice translation by $\r$. Let $\b_1, \b_2, \b_3$ be a basis for the reciprocal lattice such that $\b_i . \a_j = 2\pi \delta_{ij}$. 

We will assume that the system  is also invariant under all the symmetry operations corresponding to some nonsymmorphic space group, $G$~\footnote{However the set of all  points at which the localized tightbinding orbitals are located will not in the cases we are interested in, form a Bravais lattice, rather  they corresponds to a Bravais lattice with a basis.}.

Consider matrix representations of the group $G^{\k}$ of $\k$~\cite{herring1942character} for which the matrix corresponding to a translation by an arbitrary lattice vector,  $\r$ has the form $e^{i\k. \r}I $ where $I$ is the identity matrix. Let $n_{\k}$ be the dimension of the lowest dimensional irreducible representation of the group  of $\k$ with this property. Suppose further that there is some $\k_s$ for which $n_{\k_s} $ is strictly greater than 1. Our aim is to show that for groups which have such a point and for which some additional conditions hold, there is a finite set of points in the phase space of solenoidal flux insertions at one of which the ground state (or any other eigenstate) is necessarily degenerate. 

Owing to a similarity in structure, the space of flux insertions is conveniently parametrized by vectors in reciprocal space. A solenoidal flux insertion of $\theta_1, \theta_2$ and $\theta_3$ through the holes of the 3 torus corresponding to the directions $\a_1, \a_2 $ and $\a_3$ may be called the flux insertions associated with the wavevector $\k = \theta_1 \b_1 + \theta_2 \b_2 + \theta_3 \b_3$.  Let $H(\k)$ be the Hamiltonian which includes the effects of flux insertion associated with $\k$. Rather than using periodic boundary conditions with $H(\k)$, we may work with the original Hamiltonian, $H$ and the generalized periodic boundary conditions associated with $\k$ which may be defined as:
 $  (T_{\a_1})^{N}\ket{\psi}= e^{i\theta_1} \ket{\psi}, (T_{\a_2})^N= e^{i\theta_2}\ket{\psi},(T_{\a_3})^{N} =e^{i\theta_3} \ket{\psi}$ for single particle states $\ket{\psi}$.

For any wavevector, $\k$, the boundary conditions associated with that wavevector  are compatible with the group, $G^{\k}$ of that wavevector. To see this, let $g$ be some element of $G^{\k}$ and $R_{g}$ a proper or improper rotation associated with $g$ such that $t_{R_{g}\r}g = gt_{\r}$ for any lattice translation, $t_{\r}$ in the space group, $G$. Then,  if $\ket{\psi} $ is some single particle state which satisfies the generalized boundary conditions associated with $\k$, and $D(g)$ the unitary operator corresponding to $g$,    
\begin{align}
T_{N\a_i}D(g) \ket{\psi} &	= D(g) T_{(R_g)^{-1}N\a_i}\ket{\psi}  \nonumber \\
    & = D(g) e^{i (R_{g}^{-1} N\a_i).t\q_{g} } \ket{\psi} \nonumber \\
            & = e^{i ( N\a_i).t\q_{g} } D(g) \ket{\psi} 
\end{align}
 The last equality for $i=1,2,3$ implies that $D(g)\ket{\psi}$ also satisfies the boundary conditions associated with $\k$. 
 
 Now, if we define the twist operators, $U(\k)$: 
\begin{align}
U(\k)= e^{i {1\over N}\sum_{\r}(\k.\r )n_{\r}},
\end{align}¥

 where $n_\r$ is the fermion number operator,  and if $ \ket{\psi}$ is an eigenstate of $H(\k)$ with energy $E$ (with periodic boundary conditions), then $U(\k) \ket{\psi}$ is an eigenstate of $H$ with the same energy $E$ with the generalized periodic boundary conditions associated with $\k$. The twist operator does not commute with  translations and other space group operations. Instead, we have for any $N_p$ particle state, $\ket{\psi}$,  
\begin{align}
 T_{\r'} U(\k) \ket{\psi}  = e^{i  (N_p/N) (\k.\r') }U(\k)T_{\r'} \ket{\psi} \label{tu}
\end{align}
and if $D(g)$ is the unitary matrix corresponding to a space group operation $g$ consisting of a (proper or improper) rotation  $R$ followed by a translation by $\tau$ and if $R\k = \k$, then  
\begin{align}
 D(g) U(\k) \ket{\psi}  = e^{i  (N_p/N) (\k.\tau) }U(\k)D(g) \ket{\psi}  \label{du}
\end{align}

Suppose, further, that there is a set of integers, $n$ such that $n(\k_{s} - \k_{0})$ is a reciprocal lattice vector for any $\k_{0}$ such that the group, $G^{\k}$ of $\k=\k_{0}$ is the entire space group. Let $p$ be the smallest member of this set and let $\k'=\k_{s}-\k_{0}$. Suppose now that the ground state $\ket{\phi}$ of the Hamiltonian, $H$ with no fluxes has a momentum, $\k_{0}$ such that the group of $\k_{0}$ is the full space group, $G$. From our previous discussion, we know that there must be a continuous band of states, one for each $\k$ with the same momentum $\k_{0}$. Let  $\ket{\phi(\k)}$ denote the state which belongs to this band and is an eigenstate of $H(\k)$. 
 
 For a system where $N$ is chosen to be coprime with $p$,  consider the  set of  states $ \{U(\k') \ket{\phi(\k')}, U(2\k')\ket{\phi(2\k')} \cdots,
    U(p\k')\ket{\phi(p\k')}\} $. From Eq.~\eqref{tu}, we deduce that the set of momenta of these states consists of the wavevectors $\k_{0}+m\nu N^{2} \k'$ for  $m =1, 2, ...p $.   If $\nu$ is also coprime to $p$, then the set must contain the wavevector $\k_{s}$. There must therefore be an $0<l\le p$ such that $\k_{0} + l\nu N^{2} \k'= \k_{s}$. Then, since $n_{\k_{s}} >1$, this statement is equivalent to the statement that  there must be some $0<l\le p$ for which the Hamiltonian with boundary conditions associated with $l\k'$ or equivalently, the Hamiltonian, $H(l\k')$ must necessarily have a degeneracy. If the group of $\k_{0}$ (the momentum of the ground state) is not $G$, then for some space group element, $g$, $D(g)\ket{\phi}$ must also be an eigenstate of $H$ and has a different crystal momentum and is hence orthogonal to $\ket{\phi}$. Thus in this case there is a degeneracy of the Hamiltonian $H(l\k)$ for $l=p$. These arguments establish many-body band touchings in more than a hundred nonsymmorphic space groups including the space group of the diamond lattice, and the two exceptional nonsymmorphic space groups which do not have a nonsymmorphic element.

   We now present another set of arguments based on the monodromy of the irreducible representations of nonsymmorphic space groups~\cite{herring1942character,PhysRevB.59.5998}.     An element $g$ of $G$ is called nonsymmorphic if there is no lattice translation $t_{\r}$ such that $gt_{\r}$ leaves some point in space fixed.  For any  nonsymmorphic element $g$ of the space group $G$, there is some smallest integer $n_g$ such that $ g^{n_g} $ is  a translation by some lattice vector, $\r_{g}$.   Let $\k_g $ be the smallest reciprocal lattice vector in the direction of $\r_g$. Let  
    $G'_g$ be the subgroup of the space group generated by the primitive translations and $g$. 
     Since $G'_{g}$ is a subgroup of the group, $G^{k}$ of $\k_{g}$, the boundary conditions associated with any vector of the form $t \k_{g}$ (where $t$ is some real number) are compatible with the symmetry operations in $G'_{g}$. Thus the Hamiltonian, $H(t\k_{g})$ with solenoidal fluxes associated with $t\k_{g}$ is also invariant under these symmetry operations.

   Let $\Upsilon$ be the set of crystal momenta $\k$ such that  $G'_{g}$ is a subgroup of the group $G^{k}$ of $\k$.  
 Let $\ket{\phi}$ be some state in the ground state manifold of the system with no fluxes,  with  a crystal momentum $\k_{0} \in \Upsilon$. We can then always choose $\ket{\phi}$ such that  it is associated with  a one dimensional irreducible representation, $D^{\rho}_{\k_{0}}$ of $G'_{g}$. 
By continuity, for all $t$, there must then be an eigenstate $\ket{\phi(t)}$ of $H(t\k_{g})$ with the same momentum $\k_{0}$ and associated with the same irreducible representation,
$D^{\rho}_{\k_{0}}$ of $G'_{g}$.  If $G' _{g}$ is a subgroup of $G^{k}$ of $\k$, then there are $n_{g}$ inequivalent one dimensional irreducible representations of $G'_{g}$ associated with $\k$:
\begin{align}
 D^{p}_{\k}(g) = e^{i (\k. \r_g/n_g+ p 2\pi/n_g)}
\end{align}
 where $p$ is an integer which takes the values $1, 2 \cdots n_{g}$.

 As a consequence of Eqs.~\eqref{tu} and~\eqref{du},  the state $U(t\k_{g})\ket{\phi(t)}$ which is an eigenstate of the Hamiltonian with the boundary conditions associated with $t\k_{g}$
  has a momentum $ f(t)=\k_{0} + \nu N^{2}t \k_{g}$ and transforms per the irreducible representation, $D^{p}_{f(t)}$.  Since $\k_{g}$ is a reciprocal lattice vector, the boundary conditions associated with $\k_{g}$ are equivalent to those associated with $0$. Further, since $f(1)$ is a wavevector equivalent to $\k_{0}$,  $D^{p}_{f(1)}$ must be the same as $D^{p'}_{\k_{0}}$ for some $0<p' \le n_{g}$. In fact, $p'$ is easily seen to be $ (p+\nu N^2 (\k_{g}.\r_{g}))\mod n_g$.  It has been shown that for every element  $g$ of the type considered here,  $e^{i(\k_{g}.\r_{g})/n_{g}}\neq 1$~\cite{Konig:ProceedingsOfTheNationalAcademyOfSciences:1999}.  This implies that there are some $\nu, N$ for which  $p' \not \equiv p \mod n$. For these values of $\nu$ and $N$, the inequivalence of the representations associated with $D^{p}_{\k_{0}}$ and $D^{p}_{f(1)}$ implies that the kets $\ket{\phi}$ and $ U(\k_{g})\ket{\phi(\k_{g})}$ are orthogonal. This in turn, means that the band associated with $\ket{\phi}$ must necessarily touch some other bands in the space of solenoidal flux insertions. 
  
 We may repeat this process of flux insertion starting at each new step with the  state $\ket{\phi'} $ generated in the previous step to produce a set of orthogonal states, 
 whose number is the order of   $ ( \nu N^{2} (\k_{g}.\r_{g})) \mod n_g$ in the group of integers modulo $n_g$. The maximum possible value of this number for a given $\nu$ and $\k_g.\r_g$  is $n_g/\beta_{g}$ where $\beta_{g}$ is the greatest common divisor of $n_g$ and $\nu (\k_g.\r_g)$. In order to generate this many orthogonal states, we need $N$ to be coprime to $n_g/\beta$. 
 
For each of the nonsymmorphic elements $g$ of $G$~\footnote{It is sufficient to consider one element from each translation coset $g\mathcal{T}$ where $\mathcal{T}$ is the invariant subgroup of lattice translations of $G$.} we can use the same technique to generate connected orthogonal states. With an  $N$ which is coprime to the set of all $n_g/\beta_{g}$, we get the maximum possible lower bound on the number of orthogonal states that can be obtained in this way, which is the lowest common multiple of all of the numbers $n_{g}/\beta_{g}$. 

If the momentum $\k_{0}$ of $\ket{\phi}$ is not in $\Upsilon$, let $q$ be the smallest integer such that $g^{q} \in G^{\k_{0}}$. Then the states $\ket{\phi'_{a}}=g^{a}\ket{\phi}$ for $ 0\le a <q$ are orthogonal and necessarily degenerate. For each of these $q$ states, $\ket{\phi'_{a}}$, we can, in general, generate additional orthogonal states  which  correspond to  the element $g'=g^{q}$. The number of such states is,  using the formula derived above, $ n_{g}/ q \gamma $ where $\gamma$ is the greatest common divisor of $ n_{g}/q$ and $\nu(\k_{g}.\r_{g}) $.  This number is bounded below by $n_{g}/q \beta_{g}$, thus the total number of connected orthogonal states associated with $g$ is greater than or equal to $n_{g}/\beta$ and our previous bound on the total number of orthogonal states still holds. 

Let us now consider a system of particles with spin. We assume the system is invariant under a set of unitary transformations corresponding to the double group of some space group, $G$.  We note that all the representations of the double group of $G$ also satisfy Eqs.~\eqref{tu} and~\eqref{du}. When $\nu (2S) N^{3}$ is even, where $S$ is the spin of the particles, both our previous arguments which establish connectivity of bands apply in entirety since all eigenstate  manifolds can be decomposed into invariant subspaces associated with irreducible representations of the single group. When $\nu (2S) N^{3}$ is odd, all the eigenstate manifolds can  be decomposed into invariant subspaces of the additional irreducible representations of the double groups. 

 We are thus led to consider the additional irreducible representations of  the double group of $G ^{\k}$ such that the matrix corresponding to a translation by any lattice vector $\r$ has the form $e^{i \k.\r} I$ with $I$ being the identity matrix. Suppose there is some point $\k'_{s}$ where the minimum dimension of the additional irreducible representations of the above form of the double group is greater than $1$~\footnote{These points need not coincide with the points $\k_{s}$ associated with the single group}. We assume that, in addition, for each $\k_{0}$ such that the group, $G^{k}$ of $\k_{0}$ is the entire space group, there is some smallest integer $p$ such that $p(\k'_{s}- \k_{0})$ is a reciprocal lattice vector. The remainder of our first band connectivity argument goes through unchanged on substituting $\k_{s}$ by $\k'_{s}$.  Thus, for appropriate values of $\nu$ and $N$, we may  deduce that there is a finite set of points  in the space of flux insertions at one of which the energy bands must necessarily touch. From the tables of irreducible representations of space groups, we find that these arguments apply to a large class  of space groups~(more than a hundred). 
 
   To apply our second argument (when $\nu (2S)N^{3}$ is odd), we note that the additional irreducible representations of the double group of $G'_{g}$ which involve a single wavevector $\k_{1} $, such that $G'_{g}$ is a subgroup of the double group of $\k_{1}$, are of the form : 
   \begin{align}
   D^{p}_{\k_1}(g) = e^{i (\k_1. \r_g/n_g+ (2p+1)\pi/n_g)}
   \end{align}
   for $  0<p \le n_{g}$. Noting that these irreducible representations have the same monodromy properties as the irreducible representations of the single groups, we conclude that the number of connected orthogonal states that can be generated through flux insertion remains the same as in the case of particles without spin, for the same values of $\nu$ and $N$ for the same operation, $g$.

  If the system is invariant under just the space group transformations  of some group $G$ without the accompanying rotations in spin space~\footnote{This may happen for instance, if the system is invariant under arbitrary $SU(2)$ rotations in addition to the space group transformations.}, it can in some cases be mapped to a problem with a different filling fraction. Eigenstates of spin $1/2$ particles in a system with total $S_{z}$ and particle number conservation can be mapped to a two species bosonic model which  gives a filling factor of $1/2 - m$ per species. A quantum spin system with total conserved magnetization, $ S^{z} _{\textrm{tot}}$ can be mapped to an interacting boson system with $S-m$ particles per spin, where $S$ is the spin quantum number and $m$ is the average magnetization per spin.  Then, in the presence of space group symmetries, a fully gapped phase is not possible even for certain integer values of $n(S-m)$ where $n$ is the number of particles per unit cell. Any plateaus at the corresponding values of magnetization, if they exist, must be ``exceptional''~\cite{PhysRevLett.78.1984,oshikawa2000commensurability}.

   What we have so far shown is that under certain conditions, the energy bands of the many-body systems with integer filling fractions in the phase space of solenoidal fluxes must necessarily touch. This however necessarily  means that for these systems, with some additional constraints on the locality of the Hamiltonian, there are low lying energy states in the appropriate thermodynamic limit using the same arguments that were used by Oshikawa and Hastings for the fractional filling case~\cite{oshikawa2000commensurability,hastings2004lieb,Hastings:Epl:2007}. Essentially, a gap to local excitations in a system with interactions that are sufficiently local ensures an insensitivity to boundary conditions. This insensitivity to boundary conditions can be combined with our arguments to show that for all the nonsymmorphic space groups there is necessarily at least one (and in many cases, more than one) low lying excitation at zero flux whose gap from the ground state can be bounded by a number that goes to zero in the thermodynamic limit~\cite{Hastings:Epl:2007}. The vanishing of the matrix elements of all local operators between the different low lying states would then imply the existence of topological order. Failing this, the system must have some sort of symmetry breaking. 
  
    While our discussion has been  in 3d, the same arguments can also  be applied (with similar restrictions) to nonsymmorphic crystals in 2d to show that  for certain integer filling fractions and appropriate system sizes, there are necessarily low lying excitations. Extensions to higher dimensions  is left for future work. 
  
  I thank John Chalker and Ashvin Vishwanath for useful and stimulating discussions.~Particular thanks are due to Siddharth Parameswaran for numerous illuminating discussions, especially  on some  previous work and on the theory of space groups, and for comparison of results~\footnote{We  have learned that parallel work by Parameswaran et al. related to these topics has recently been posted on the arXiv at arXiv:1212.0557v2.}.

\end{document}